\documentclass[12pt]{article}
\usepackage{amssymb}
\usepackage{amsfonts}       
\usepackage{graphicx}
\usepackage{ams math}
\usepackage{tikz}
\usepackage{hyperref}
\usepackage[siunitx, RPvoltages]{circuitikz}
\begin{document}
	\title{On Logic Gates with Complex Numbers}
	\author{  M. W.   AlMasri \thanks{mwalmasri2003@gmail.com, almasri@sjtu.edu.cn} \\ \small 
	Wilczek Quantum Center, School of Physics and Astronomy, Shanghai Jiao Tong University,\\ 
	\small 800 Dongchuan Road, Minhang District, Shanghai, 200240, China}
	\maketitle
	\begin{abstract}
Logic gates can be written in terms of complex differential operators, where the inputs and outputs are holomorphic functions with several variables. Using the polar representation of complex numbers, we arrive at an immediate connection between the oscillatory behavior of the system and logic gates.  We discuss the universality of this formalism in a variety of computing systems. 
	\end{abstract}
	\section{Introduction}
 In his book {\it Calculating Space}, Zuse attempted to describe the laws of physics in terms of a discrete cellular automata model \cite{Zuse}. This was one of the earliest attempts in connecting physical laws with computational models. More details on the connection between cellular automata and physics can be found in the book \cite{automata}. From hardware perspective, computers are physical systems: the laws of physics dictate what they can and cannot do. Concretely, the speed with which a physical device can process information is limited solely by its energy, and the amount of information that it can process is limited by the number of degrees of freedom it possesses. Therefore, there must be an intimate relationship between the dynamical evolution of  physical states and their behavior regarding processing and storing information \cite{Landauer1, Landauer,Feynman,Bennett,Lloyd,Parrondo}.\vskip 5mm
 
 Dynamical systems such as physical processes can be mapped into a computational model.   This  model consists of a computing machine $M$ connected to an environment $E$. At each time $t$, we have input channel(s) $S(t)$ (for stimulus) and output channel(s) $R(t)$ (for response) \cite{Minsky}. To describe how the computing machine $M$ works, we need to specify the outputs as a function of the inputs. Since the transmission of signals requires some time, the output must be at a time later than the time of the input channel. We assume the machine response $R(r+1)$ because of  $S(t)$ to depend on the  state of the machine  $Q(t) $ \cite{Minsky}
	 \begin{eqnarray}\label{R}
	 	R(t+1)= F(Q(t),S(t))
	 \end{eqnarray}
 and the state of the machine $M$ depends on the stimulus $S(t)$ and the state of the machine at previous time, 
 \begin{eqnarray}\label{Q}
 	Q(t+1)= G(Q(t),S(t)). 
 \end{eqnarray}
 
For example, the memristive systems can be  defined as dynamical systems by \cite{Chua}
\begin{eqnarray}\label{memristive1}
	\dot{x}= f(x,u,t),\\
	y= g(x,u,t) u,\label{memristive2}
\end{eqnarray}
where $u$ and $y$ are the input and output of the system, and $x$ denotes the state of the system. The connection between \ref{memristive1}, \ref{memristive2} and \ref{R}, \ref{Q} is achieved by assigning $R(t+1)=y(t+1)$, $Q(t+1)=x(t+1)$, and $u=S(t)$. Therefore, for each value of $t$, we can realize the memristive system as a computational model. The same procedure can be applied to any other dynamical system.\vskip 5mm
According to quantum mechanics, subatomic particles behave like waves propagating in spacetime. The wave-particle duality was proposed by de Broglie  \cite{waves}. The de Broglie wavelength $\lambda$ associated with a particle is
\begin{eqnarray}
	\lambda=\frac{h}{p}, 
\end{eqnarray}
where $h$ is Planck's constant and $p$ is the particle's momentum. In the Davisson-Germer experiment, electrons scattered by the surface of a crystal of nickel metal produced a diffraction pattern similar to waves\footnote{C. Davisson  and L. H. Germer, \emph{Phys. Rev.} {\bf 30} (1927) 705}. Interestingly, changing diffraction patterns, as we shall see in the incoming sections, are intimately connected to the operations of logic gates. Physical quantities such as the position and momentum are expressed in terms of periodic functions in time. From the very basic level, varying complex numbers  with  respect to $t$ give wave-like behavior since  $z_{i}(t)=|z_{i}|e^{i\theta(t)}= |z_{i}|(\cos\theta(t)+i\sin\theta(t))$.  This suggests the possibility of building logical components based on the distinct vibrational states of particles. \vskip 5mm
In this article, we formulate logic gates in terms of holomorphic variables. This was inspired by the holomorphic representation of quantum mechanics. However, this formulation can be used for any dynamical system described by a set of holomorphic functions. The state functions in quantum systems have a different pattern comparing with classical systems. The correspondence between oscillatory behavior and logic gates computation is assured mathematically by writing logic gates in terms of complex differential operators acting on holomorphic functions. 
Oscillatory systems are omnipresent in universe. Some notable examples are the attractor cycle oscillators, rotors, neural oscillation patterns in the brain, synchronized chaos, and Josephson junctions. Other examples in biology and physics  can be read from \cite{Winfree,Strogatz}. 
\section{Holomorphic Representation of Logic Gates}
	
	In this section, we formulate  logic gates as differential operators written in terms of complex variables and their partial derivatives. Such representation is possible both in classical and quantum regimes. To see this, consider a hypothetical material which allows red light to pass through it and absorbs black light. In this case, if we assign the variable $z_{1}$ for the red light and $z_{2}$ for the black light. We may represent the action of the material by the differential operator $\frac{\partial}{\partial z_{2}}$. Generally, it is possible to find equivalent expressions for  propositional logic in terms of holomorphic variables and their partial derivatives.

 \subsection{Holomorphic representation of quantum logic gates}
 The states of a quantum system with $n$ degrees of freedom are usually described by functions either in the configuration space $(q_{1},q_{2},\dots q_{n})$ or in the momentum space $(p_{1},p_{2},\dots p_{n})$. The complex combinations of these variables have proven to be effective both at quantum and classical levels. Moreover, they appear in the study of harmonic oscillator and Bose particle field theories as the creation and annihilation operators. \vskip 5mm
 We define the complex conjugate operators as
 \begin{eqnarray}
 	\eta_{k}=2^{-1/2} \left(q_{k}-i p_{k}\right), \\
 	\xi_{k}= 2^{-1/2}\left(q_{k}+ip_{k}\right).
 \end{eqnarray} 
 If $q_{k},p_{k}$ are self-adjoint operators satisfying the canonical commutation relations (with Planck constant $\hbar=1$)
 \begin{eqnarray}
 	[q_{k},p_{l}]=i\delta_{kl}, \; [q_{k},q_{l}]=[p_{k},p_{l}]=0,
 \end{eqnarray}
Then it follows that 
\begin{eqnarray}
	\xi_{k}=\eta^{\star}_{k}, \; \eta_{k}=\xi_{k}^{\star},\;  [\xi_{k},\eta_{l}]=\delta_{kl}, \; [\xi_{k},\xi_{l}]=[\eta_{k},\eta_{l}]=0.
\end{eqnarray}
In 1928, Fock  introduced the operator solution $\xi_{k}=\partial/\partial \eta_{k}$ of the commutation relation $[\xi_{k},\eta_{k}]=1$  in analogy with $p_{k}=-i\partial/\partial q_{k}$ of the relation $[q_{k},p_{k}]=i$ and applied it to quantum field theory\footnote{V. A. Fock, \emph{Verallgemeinerung und Loesungder Diracschen statistischen Gleichung}, Z.Phys. {\bf 49},339 (1928).}. \vskip 5mm 
Consider a quantum  harmonic oscillator with a single-mode described in terms of the  coordinates $q,p$ in the phase space. The creation and annihilation operators are 
\begin{eqnarray}
\eta=2^{-1/2} \left(q-i p\right), \\
\xi= 2^{-1/2}\left(q+ip\right).
\end{eqnarray}
with canonical commutation relation $[\xi,\eta]=1$. The Hamiltonian operator is 
\begin{eqnarray}
	H= \eta \frac{\partial }{\partial \eta}+\frac{1}{2},
\end{eqnarray}
with energy eigenvectors equal to $\frac{\eta^{n}}{\sqrt{n!}}$. 
 \vskip 5mm 
It is convenient as it will appear later in this work to represent  spin operators in terms of uncoupled quantum oscillators. Here, we use the Jordan-Schwinger map of the $\mathfrak{su}(2)$ algebra to deal with particles with arbitrary spin number by means of creation and annihilation operators. For two uncoupled quantum oscillators with annihilation operators $a$ and $b$, the spin operators read, 
 \begin{eqnarray}\label{spin1}
 	\sigma_{x}=\left(a^{\dagger}b+b^{\dagger}a\right), \\
 	\sigma_{y}= -i\left(a^{\dagger}b-b^{\dagger}a\right),\label{spin2}\\
 	\sigma_{z}=\left(a^{\dagger}a-b^{\dagger}b\right).\label{spin3}
 \end{eqnarray}
The canonical commutation relations between the spin operators are 
\begin{eqnarray}
	[\sigma_{i},\sigma_{j}]=i\;\epsilon_{ijk}\;\sigma_{k}, \;\;\; [\sigma_{i},\sigma^{2}]=0, 
\end{eqnarray}
where $i=x,y,z$. The squared spin operator reads \begin{equation}
\sigma^{2}=\frac{N}{2}\left(\frac{N}{2}+1\right)
\end{equation}
where $N=N_{1}+N_{2}$  is the total number operator with $ N_{1}=a^{\dagger}a$ and $N_{2}=b^{\dagger}b$. 
  Normalized states with occupation numbers $n_{1}$ and $n_{2}$ can be obtained by applying the creation operators on the vacuum state i.e. $|0,0\rangle$,  
\begin{eqnarray}
	|n_{1},n_{2}\rangle= \frac{(a^{\dagger})^{n_{1}}}{\sqrt{n_{1}!}}\frac{(b^{\dagger})^{n_{2}}}{\sqrt{n_{2}!}}|0,0\rangle.
\end{eqnarray} 
With $j=\frac{n_{1}+n_{2}}{2}$ and $m=\frac{n_{1}-n_{2}}{2}$ runs from $-j$ to $j$ in integer steps. \vskip 5mm
In Bargmann representation, the spin operators \ref{spin1}, \ref{spin2} and \ref{spin3} read \cite{almasri1,almasri2,almasri3}
\begin{eqnarray}
\sigma_{x}=\left(z_{1}\frac{\partial}{\partial z_{2}}+z_{2}\frac{\partial }{\partial z_{1}}\right), \\
\sigma_{y}= -i\left(z_{1}\frac{\partial}{\partial z_{2}}-z_{2}\frac{\partial }{\partial z_{1}}\right),\\
\sigma_{z}=\left(z_{1}\frac{\partial }{\partial z_{1}}-z_{2}\frac{\partial }{\partial z_{2}}\right).
\end{eqnarray}
In this case, the normalized states are 
\begin{eqnarray}\label{normalized}
|n_{1},n_{2}\rangle=\frac{z_{1}^{n_{1}}}{\sqrt{n_{1}!}} \frac{z_{2}^{n_{2}}}{\sqrt{n_{2}!}}\equiv f_{n_{1}n_{2}}
\end{eqnarray} 
since 
\begin{align}
	\langle m_{1},m_{2}|n_{1},n_{2}\rangle=\frac{1}{\pi^{2}}\int_{\mathbb{C}^{2}}\overline{f_{m_{1}m_{2}}}\; f_{n_{1}n_{2}} \; e^{-|z_{1}|^{2}-|z_{2}|^{2}}\; dz_{1} \; dz_{2}=  \delta_{m_{1},n_{1}} \delta_{m_{2},n_{2}}.
\end{align}

 The fundamental logic gates that operate between two holomorphic functions are  the Pauli gates plus the identity gate $(I,  \vec{\sigma})$\cite{Nielsen}. Any given two-input gate can be written in terms of the identity operator and the Pauli vector. 
	  Pauli gates are defined as complex differential operators that act on holomorphic functions with two variables
	 \begin{eqnarray}
	 	X=\mathrm{NOT}= \sigma_{x}=\left(z_{1}\frac{\partial}{\partial z_{2}}+z_{2}\frac{\partial }{\partial z_{1}}\right),\\
	 	Y=\sigma_{y}=-i\left(z_{1}\frac{\partial}{\partial z_{2}}-z_{2}\frac{\partial }{\partial z_{1}}\right),\\
	 	Z=\sigma_{z}=\left(z_{1}\frac{\partial }{\partial z_{1}}-z_{2}\frac{\partial }{\partial z_{2}}\right).
	 \end{eqnarray}
For spin-1/2 case, we have two distinct state functions: $|1,0\rangle=z_{1}$ which corresponds to spin-up ($m=1/2$), and $|0,1\rangle=z_{2}$ for spin-down ($m=-1/2$).
 The identity gate in the case of a two-variable function is 
 \begin{eqnarray}
 	I=\left(z_{1}\frac{\partial }{\partial z_{1}}+z_{2}\frac{\partial }{\partial z_{2}}\right)
 \end{eqnarray}
For the $N$-dimensional case, the identity operator in the Bargmann space reads , 
\begin{eqnarray}
	I= \sum_{i=1}^{N} z_{i}\frac{\partial }{\partial z_{i}}. 
\end{eqnarray}

\vskip 5mm
The  Hadamard or Walsh-Hadamard gate acts on holomorphic functions with two-variables in the following way 
\begin{eqnarray}
H=\frac{1}{\sqrt{2}}\left(z_{1}\frac{\partial }{\partial z_{1}}+z_{1}\frac{\partial }{\partial z_{2}}+z_{2}\frac{\partial }{\partial z_{1}}-z_{2}\frac{\partial }{\partial z_{2}}\right)
\end{eqnarray}


The controlled-Not gate acts on holomorphic  functions with four-variables,
it consists of an identity operator with respect to a two-variables  $z_{1},z_{2}$ and a Not-gate operator between the remaining two,
\begin{eqnarray}
	\mathrm{CNOT}=\left(z_{1}\frac{\partial}{\partial z_{1}}+z_{2}\frac{\partial }{\partial z_{2}}+z_{3}\frac{\partial}{\partial z_{4}}+z_{4}\frac{\partial }{\partial z_{3}}\right). 
\end{eqnarray}
\vskip 5mm 
The Swap , Toffoli and Fredkin gates are
\begin{eqnarray}
	\mathrm{Swap}=\left(z_{1}\frac{\partial }{\partial z_{1}}+z_{2}\frac{\partial }{\partial z_{3}}+z_{3}\frac{\partial}{\partial z_{2}}+z_{4}\frac{\partial }{\partial z_{4}}\right),
\end{eqnarray}

\begin{align}
	\mathrm{Toffoli}=\left(z_{1}\frac{\partial}{\partial z_{1}}+z_{2}\frac{\partial}{\partial z_{2}}+z_{3}\frac{\partial}{\partial z_{3}}+z_{4}\frac{\partial}{\partial z_{4}}+z_{5}\frac{\partial }{\partial z_{5}}+z_{6}\frac{\partial }{\partial z_{6}}+z_{7}\frac{\partial}{\partial z_{8}}+z_{8}\frac{\partial }{\partial z_{7}}\right), 
\end{align}

\begin{align}
	\mathrm{Fredkin}=\left(z_{1}\frac{\partial}{\partial z_{1}}+z_{2}\frac{\partial }{\partial z_{2}}+z_{3}\frac{\partial }{\partial z_{3}}+z_{4} \frac{\partial }{\partial z_{4}}+ z_{5}\frac{\partial}{\partial z_{7}}+z_{7}\frac{\partial }{\partial z_{5}}+z_{8}\frac{\partial }{\partial z_{8}}\right). 
\end{align}
Other possible logic gates can be written as complex differential operators in a way similar to the procedure we followed in writing the previous gates. If the input function is of the form \ref{normalized}, the outputs are again normalized states. The expectation values of $\langle n_{1},n_{2}|L|n_{1},n_{2}\rangle$, where the differential operator $L$ is any logic gate operator such as $X,Y,Z,H$ etc., and the inner product is taken over the Bargmann space, are identical to the expectation values of $\langle \psi | \hat{L}|\psi\rangle$ with a wave function $\psi$, gate operator $\hat{L}$ and an inner product taken over the vector Hilbert space. Moreover, considering the inputs to be any holomorphic functions is also possible. However, in this case, the normalization condition may be lost, but it could be recovered by dividing with a suitable normalization constant. Below, we give a general procedure for obtaining the differential and integral representations of logic gates in the two-dimensional Bargmann space. The formal definition of the Bargmann spaces and its related properties are mentioned in the appendix of this manuscript.\vskip 5mm
\subsection{Differential operator representation of arbitrary logic gates} 
Let $M$ be any $2\times 2$ matrix in $\mathbb{C}^{2}$, the mapping to analytical form in Bargmann space is
\begin{equation}
M= \begin{pmatrix}
a &b\\
c &d
\end{pmatrix}\rightarrow {\bf M}(z_{1},z_{2})= a\;z_{1}\frac{\partial}{\partial z_{1}}+ b\; z_{1} \frac{\partial }{\partial z_{2}} + c\; z_{2} \frac{\partial}{\partial z_{1}} + d\;  z_{2} \frac{\partial} {\partial z_{2}}, 
\end{equation}
where $a,b,c$ and $d \in \mathbb{C}$. \vskip 2mm 
Let $f(z_{1},z_{2})$ be holomorphic function of two variables $z_{1}$ and $z_{2}$. The expansion of $f(z_{1},z_{2})$ is 
\begin{eqnarray}
f(z_{1},z_{2})=\sum_{n,m=0}^{\infty} C_{nm} z_{1}^{n}z_{2}^{m}
\end{eqnarray}
where the coefficients $C_{nm}$ are given by 
\begin{eqnarray}
C_{nm}=\frac{1}{n!m!}f^{(n)}(0)f^{(m)}(0)
\end{eqnarray}
The inner product of $f(z_{1},z_{2})$ with itself is 
\begin{eqnarray}
\langle f(z_{1},z_{2})|f(z_{1},z_{2})\rangle = \sum_{n^{\prime},m^{\prime}=0}^{\infty} \sum_{n,m=0}^{\infty}\;  C^{\star}_{n^{\prime}m^{\prime}}\; C_{nm}\;  n!\;  m!\;  \delta_{n^{\prime},n} \delta_{m^{\prime},m}
\end{eqnarray}
The action of ${\bf M}(z_{1}, z_{2})$ on $f(z_{1},z_{2})$ is 
\begin{align}
{\bf M}(z_{1},z_{2})f(z_{1},z_{2})= a \;n \;f(z_{1},z_{2})+ d\; m\; f(z_{1},z_{2})\\ \nonumber+ b \;m \; \sum_{n,m=0}^{\infty} C_{nm} z_{1}^{n+1} z_{2}^{m-1} + c\; n \; \sum^{\infty}_{n,m=0} z_{1}^{n-1} z_{2}^{m+1}
\end{align} 
The generalization of the above recipe to higher-order logic gates could  be done analogously in a straightforward manner. 
\subsection{ Integral representation of logic gates}
Let $f(z)$ be an analytic function in simply-connected domain $D$, and let $\Gamma$ be positively-oriented loop in $D$. The integral representation of the derivative at point $z_{0}$ (any point in the interior of $\Gamma$) is  \cite{Ahlfors}
\begin{eqnarray}
f^{(n)}(z_{0})=\frac{n!}{2\pi i}\oint_{\Gamma} \frac{f(z)}{(z-z_{0})^{n+1}}\; dz, 
\end{eqnarray}
which is known as the generalized Cauchy formula. Consequently, the integral representation of the $X$-gate for example reads 
\begin{eqnarray}
X f(z_{1},z_{2})= \frac{z_{1}}{2\pi i} \oint_{\Gamma_{2}} \frac{f(z_{1},\xi_{2})}{(\xi_{2}-z_{2})^{2}} d\xi_{2}+\frac{z_{2}}{2\pi i} \oint_{\Gamma_{1}}\frac{f(\xi_{1},z_{2})}{(\xi_{1}-z_{1})^{2}} d\xi_{1}, 
\end{eqnarray}
Similarly, we can find the integral representation for the $Y,Z$ and $I$ gates and any other logic gate. 

	 \section{The Computational Model }

Let $|{\bf f}\rangle=\left(|f_{1}\rangle,|f_{2}\rangle,\dots |f_{N}\rangle\right)$ be a vector of analytic complex functions with several variables ( the stimulus $S$), and let $M ({\bf L})$ be a computing machine with ${\bf L}$ being $N\times N$  complex differential operators diagonal matrix . The computation process $\mathfrak{C}$ is described by the action of ${\bf L}$ on the inputs $|{\bf f}\rangle$: 
\begin{equation} 
\mathfrak{C}: |{\bf f}\rangle \rightarrow {\bf L}|{\bf f}\rangle, 
	\end{equation}
and the expectation values of the outputs (the response $R$) are 
\begin{eqnarray}
	\frac{\langle {\bf f }|{\bf L}|{\bf f}\rangle}{\langle {\bf f}|{\bf f}\rangle}
\end{eqnarray}
where we divided the last expression by the normalization constant $\langle {\bf f}|{\bf f}\rangle$ since the inputs are arbitrary complex functions. Explicitly, $\langle {\bf f}|{\bf L}|{\bf f}\rangle= \langle f_{1}|L_{1}|f_{1}\rangle+ \langle f_{2}|L_{2}|f_{2}\rangle+\dots +\langle f_{N}|L_{N}|f_{N}\rangle$ and $\langle {\bf f}|{\bf f}\rangle= \langle f_{1}|f_{1}\rangle+ \langle f_{2}|f_{2}\rangle+\dots +\langle f_{N}|f_{N}\rangle$.   The Figure 1 shows the computational model described here.  If both the inputs and outputs are discrete and not correlated between themselves, i.e. cannot be written in superposition states like quantum states, then the computation is classical. Quantum computations are characterized by the possibility of having superposition states during the computation process.  The machine $M({\bf L})$  could be in principle any dynamical system such as physical objects, stock prices, ecological models, and in principle any changing system described by a set of differential equations.   \vskip 5mm
\begin{figure}
	\centering
	\begin{tikzpicture}
		\node[draw, rectangle, minimum width = 3 cm, minimum height = 2 cm] (fl) at (0,0) {$M({\bf L})$};
		\node[above] at (fl.north) {};
		\draw[-] (fl) -- node[above]{$|{\bf f}\rangle$} node[below]{Inputs} ++(-4,0);
		\draw[-] (fl) -- node[above]{${\bf L}|{\bf f}\rangle=|{\bf f}^{\prime}\rangle$} node[below]{Outputs} ++(4,0);
	\end{tikzpicture}
\caption{The computational model consists of a computing machine $M$ described by $N\times N$  complex differential operators diagonal matrix  ${\bf L}$ with inputs $|{\bf f}\rangle $ and  outputs ${\bf L}|{\bf f}\rangle$.   }
\end{figure}

To quantify the computation process, we compute the change of entropy \cite{Shannon,Cover} 
\begin{eqnarray}
	\Delta S= S_{\mathrm{out}}-S_{\mathrm{in}}= -\sum_{i} p^{\mathrm{out}}_{i}\log p^{\mathrm{out}}_{i}+\sum_{i} p^{\mathrm{in}}_{i} \log p_{i}^{\mathrm{in}}
\end{eqnarray}
where $p_{i}^{\mathrm{in}}= \frac{|f_{i}\rangle \langle f_{i}|}{\langle f_{i} |f_{i} \rangle} $ and $p^{\mathrm{out}}_{i}= \frac{|f^{\prime}_{i}\rangle \langle f^{\prime}_{i}|}{\langle f_{i}^{\prime}|f_{i}^{\prime}\rangle}$ with $|f_{i}^{\prime}\rangle =L_{i}|f_{i}\rangle$. \vskip 5mm

The relative entropy (Kullback–Leibler divergence) is 
\begin{eqnarray}
	D(P|| Q)= \sum_{i}p_{i} \log(\frac{p_{i}}{q_{i}})= -\sum_{i} \frac{|f_{i}\rangle \langle f_{i}|}{\langle f_{i}| f_{i}\rangle }\log \frac{|g_{i}\rangle \langle g_{i}|}{\langle g_{i}|g_{i}\rangle }- S(p_{i})
\end{eqnarray}
	with $p_{i}=\frac{|f_{i}\rangle \langle f_{i}|}{\langle  f_{i}| f_{i} \rangle}$, $q_{i}=\frac{|g_{i}\rangle \langle g_{i}|}{\langle g_{i}| g_{i}\rangle}$ and $S(p_{i})=-\sum_{i} p_{i}\log p_{i}$. Other quantities such as the mutual information can be determined in terms of holomorphic functions \cite{Cover}.

	 \section{Applications } 
	 All physical systems process information. In this section, we study in some sort of detail the computation mechanisms in a few basic systems.  However, it is obvious that our construction works for all physical systems in nature. \vskip 3mm  
	 \subsection{ Coupled oscillators} Simple harmonic motion is omnipresent in many  mechanical and electrical systems. For example,  electrical circuits with inductance $L$ connected across a capacitance $C$ carrying charge $Q$ obey the equation  
	  \cite{Pain}
	  \begin{eqnarray}
	  	L \ddot{Q}+\frac{Q}{C}=0
	  \end{eqnarray}
  with frequency $\omega= \frac{1}{\sqrt{LC}}$. \vskip 5mm
  The general equation of motion for any simple harmonic motion with $N$ oscillators reads,
  \begin{eqnarray}
  	\ddot{{\bf x}}+\mathbf{\omega}^{2} {\bf x}=0
  \end{eqnarray}
where ${\bf x}=\left(x_{1},x_{2},\dots ,x_{N}\right)$ is the $N$-dimensional state vector and $\omega$ is the $N\times N$ frequency matrix. 
\begin{figure}\label{spring}
	\centering
	\begin{tikzpicture}[]
		\draw (0,0) -- ++(0,2) node[left, midway]{$\ell$};
		\draw (2,0) -- ++(0,2) node[right, midway]{$\ell$};
		\draw (0,0) to[cute inductor,
		inductors/coils=9, inductors/width=1.4,
		nodes width=0.1, l=$s$, 
		o-o] (2,0);
		\draw (0,-.5) node[flowarrow, label=below:$y$]{};
		\draw (2,-.5) node[flowarrow, label=below:$x$]{};
	\end{tikzpicture}
\caption{Two identical pendulums with mass $M$ connected by a spring with stiffness $s$.}
\end{figure}
To define the basic logic gates in such systems, one needs at least two coupled oscillators at each time step $t_{0}$ \footnote{On the contrary, one could apply the basic logic gates on a single quantum particle  due to the superposition property (which leads to the quantum entanglement) of these particles, which distinguishes quantum particles from classical ones. }. Consider two  identical pendulums, each having a mass $M$ suspended on a light rigid
rod of length $\ell$. The masses are connected by a light spring of stiffness $s$ as shown in Figure 2. The natural length of the spring is equal to the distance between the two masses at equilibrium.  The equation of motion are 
\begin{eqnarray}
M	\ddot{x}= -M g\;\frac{x}{\ell} - s\;(x-y), \\
M \ddot{y}=-M g\;\frac{y}{\ell}+ s\;(x-y).
\end{eqnarray}
The above set of equations could be written as 
\begin{eqnarray}\label{pendulum}
	\ddot{x} +\omega_{0}^{2}\;x = - \frac{s}{M}\;(x-y), \\ \label{pendulum1}
	\ddot{y}+ \omega_{0}^{2}\;y=\frac{s}{M}\;(x-y).
\end{eqnarray}
where $\omega_{0}=\sqrt{\frac{g}{l}}$ is the natural frequency of each pendulum. To solve \ref{pendulum} and \ref{pendulum1}, we introduce the normal coordinates 

\begin{eqnarray}
	X=x+y,\\
	Y= x-y. 
\end{eqnarray}
Then, we have the following set of equations 
\begin{eqnarray}\label{e1}
\ddot{X}+ \omega_{0}^{2} X=0,\\ \label{e2}
\ddot{Y}+ \left(\omega_{0}^{2}+ 2\frac{s}{M}\right)	Y=0. 
\end{eqnarray}
One possible solution to the above set of equations is
\begin{eqnarray}
X= A\sin(\omega_{0}t+\phi)= \alpha (z_{1}-\overline{z}_{1}),\\
Y= B\sin(\omega t+\varphi)=\beta (z_{2}- \overline{z}_{2}), 
\end{eqnarray}
where $z_{1}= e^{(i\omega_{0}t+\phi)}$, $z_{2}= e^{(i\omega t+\varphi)}$, and $\omega=\sqrt{\omega_{0}^{2}+ 2s/M}$. It is sufficient to take the complex part without its complex conjugate during our study of the logic gates for coupled oscillators. The state function of the coupled oscillator is 
\begin{eqnarray}
	|f\rangle=f(z_{1},z_{2})= \alpha \beta\;  z_{1}z_{2}
\end{eqnarray}
The action of logic gates on this state function is 
\begin{eqnarray}
	Xf=\mathrm{NOT}f= \alpha\beta \left(z^{2}_{1}+z^{2}_{2}\right), \\
	Yf= -i \alpha \beta \left(z_{1}^{2}- z_{2}^{2}\right), \\
	Zf= \alpha \beta \left(z_{1}z_{2}-z_{2}z_{1}\right)=0,\\
	If= \alpha \beta \left(z_{1}z_{2}+z_{2}z_{1}\right)=2\; \alpha\;\beta \;f, 
\end{eqnarray}
where the factor 2 in the last equation can be absorbed safely choosing suitable normalization constant. The Hadamard gate is 
\begin{eqnarray}
	Hf= \frac{\alpha \; \beta }{\sqrt{2}} \left(z_{1}z_{2}+ z_{1}^{2}+ z_{2}^{2}- z_{2}z_{1}\right)=\frac{\alpha \; \beta }{\sqrt{2}}\left(z_{1}^{2}+ z_{2}^{2}\right)=\frac{1}{\sqrt{2}}Xf
\end{eqnarray} 
Note that since our state function is classical (product state), we were able to write the action of the Hadamard gates in the previous equation in terms of the action of the $X$ gate. The inner products $\langle f|X|f\rangle=\langle f|Y|f\rangle=\langle f|H|f\rangle=0$. If one considers a more general model of higher-order oscillations,  we may use the ansatz 
\begin{eqnarray}
	f(z_{1},z_{2})=\frac{z_{1}^{n}}{\sqrt{n!}}\frac{z^{m}_{2}}{\sqrt{m!}}
\end{eqnarray}
and the expectation values of the logic gates assume a more general form. As an example, we find for the $X$-gate 
\begin{eqnarray}
	X|f\rangle= m \frac{z_{1}^{n+1}}{\sqrt{n!}} \frac{z_{2}^{m-1}}{\sqrt{m!}}+ n  \frac{z_{1}^{n-1}}{\sqrt{n!}} \frac{z_{2}^{m+1}}{\sqrt{m!}},\\
\langle f^{\prime}|X|f\rangle= m\; \delta_{n^{\prime},n+1}\delta_{m^{\prime},m-1}+n\; \delta_{m^{\prime},m+1}\delta_{n^{\prime},n-1}. 
\end{eqnarray} 
Similar relations could be found for other gates. 
 \subsection{Turing Patterns}
Turing Patterns are described by  the diffusion-reaction equations \cite{Turingbio,patterns}
\begin{eqnarray}
	\frac{\partial a}{\partial t}= D_{a} \nabla^{2}a+ f(a,b),\label{tur1}\\
	\frac{\partial b}{\partial t}=D_{b} \nabla^{2}b+g(a,b)\label{tur2},
\end{eqnarray}
where $a$ and $b$ describe the concentrations of chemicals at time $t$. The functions $f(a,b)$ and $g(a,b)$ represent the reaction terms, and $D_{a},D_{b}$ are the diffusion coefficients.  One could for example write the reaction equations using the FitzHugh–Nagumo equation. In this case, $f(a,b)= a-a^{3}-b+\alpha$ and $g(a,b)= \beta (a-b)$ where $\alpha$ and $\beta$ are constants. By solving equations \ref{tur1} and \ref{tur2}, we obtain the concentrations of chemicals as a function of position and time i.e. $\{a(x,t), b(x,t)\}$. The chemical concentrations a and b belong to different analytic Hilbert spaces since each describes one type of chemical material. Using the Segal-Bargmann transform, we may write the solutions in terms of holomorphic variables as 
\begin{eqnarray}
	\hat{a}(z_{1},t)= (\pi t)^{-1/4} \int_{\mathbb{R}^{d}} e^{\left(- z_{1}^{2}+2\sqrt{2}z_{1}\cdot x- x^{2}\right)/2t} a(x)\;dx,\\
	\hat{b}(z_{2},t)= (\pi t)^{-1/4} \int_{\mathbb{R}^{d}} e^{\left(-z_{2}^{2}+2\sqrt{2}z_{2}\cdot x-x^{2}\right)/2t} b(x)\;dx.
\end{eqnarray}
The state function of the system is 
\begin{eqnarray}
	f(z_{1},z_{2})= \hat{a}( z_{1},t)\hat{b}(z_{2},t)
\end{eqnarray}
which is a product state. Feeding the state function into logic circuit architectures built from complex differential operators defines the computational processes of the Turing patterns.
 
\subsection{Neural Networks}
We describe the mechanism for incorporating the developed formalism into the study of neural networks. We will restrict our analysis to simple neural networks. However, the generalization to other types of nets such as convolutional neural networks is possible. The neuron $j$ may be described mathematically by a set of two equations \cite{Mcculloch,Rosenblatt,Hopfield,Haykin,deep,Almasri}, 
\begin{eqnarray}
	u_{j}= \sum_{i=1}^{N}\omega_{ji}x_{i}, \\
	y_{i}= f(u_{j}+b_{j}),
\end{eqnarray} 
where $\omega_{j1},\omega_{j2},\dots ,\omega_{jN}$ are the synaptic weights, $f$ is the activation function and $b_{k}$ is the bias. The inputs are
\begin{eqnarray}\label{inputs}
	x_{1}=\langle f_{1}|L_{1}|f_{1}\rangle, x_{2}= \langle f_{2}|L_{2}|f_{2}\rangle, \dots, x_{N}=\langle f_{N}|L_{N}|f_{N}\rangle.
\end{eqnarray} 
The outputs $y_{j}$ should be written as inner products in the Bargmann space similar to the inputs in \ref{inputs}. Now, following any neural computation algorithms for weights such as  perceptrons, backpropogations one could find the convergence  learning process with minimal loss functions. 
	\section{Universal Programming Language }
Programming languages are systems of expressions mostly text-based formal languages that allow users to communicate with the machine language \cite{Knuth}.  The source code of any software written in any programming language such as LISP and C is translated at the end into binary codes when using our personal computers \cite{lisp,lisp1,c}. Binary codes are written using two truth values i.e. 0 and 1. At hardware level, they are encoded using bits which in turn can be represented uniquely using switches within the central processing unit (CPU) where ``1=On" and ``0=Off".  Binary codes  can be represented in our construction using two holomorphic functions $f(z)$ and $g(z)$  with different polynomial  degrees i.e. $deg(f)\neq deg(g)$. The simplest choice is to represent the truth value $0$ by  $z^{0}=1$ and $1$ by $z$. For multiple-valued logic in which we have more than two truth values (0 and 1), we may use $\frac{z^{N}}{\sqrt{N!}}$ to represent the truth values $N=0,1,2\dots \in \mathbb{N}$.  In these schemes, the binary code $\texttt{01101011}$ reads as $\texttt{f(z)g(z)g(z)f(z)g(z)f(z)g(z)g(z)}$ using general holomorphic functions and  $\texttt{1zz1z1zz}$ using monomials as logical states. \vskip 5mm 

A Universal Programming Language (UPL)  uses holomorphic functions for encoding logical states  and complex differential operators for the transformation between different logical states. The ordinary programming languages that run on digital computers are a special class of a more general universal programming language since any binary code is a subset from a more general coding scheme based on continuous  holomorphic functions rather than two discrete values (1 and 0). 

\vskip 5mm
Let $\mathcal{S}=\{\mathcal{S}_{1},\mathcal{S}_{2},\mathcal{S}_{3},\dots, \mathcal{S}_{N}\}$ be the system that we want to study its computational properties. Generally, we assume the system to be made of many subsystems. Each subsystem $\mathcal{S}_{i}$, where $i\in \{1,\dots ,N\}$, is described by a state function $|f_{i}\rangle$. Let  ${\bf |f\rangle}=(|f_{1}\rangle, |f_{2}\rangle,|f_{3}\rangle, \dots ,|f_{N}\rangle)$ be the joint state function for the system $\mathcal{S}$.  The general procedure to find the computational properties of $\mathcal{S}$ is determined by the following list of instructions $\mathcal{L}=\{L_{1},L_{2},L_{3},L_{4},L_{5},L_{6},L_{7},L_{8}\}$: 

\begin{itemize}
	\item $L_{1}$: Determine the state functions $\{|f_{i}\rangle\}$ of the subsystems $\{\mathcal{S}_{i}\}$ as functions of time. 
	\item $L_{2}:$ If the state functions are defined over the field of real numbers, apply the Segal-Bargmann transform for each variable such that the final answer is written in terms of holomorphic variables. \footnote{In principle, one could use any unitary integral transform from real variables onto holomorphic variables.} 
	\item $L_{3}:$ Determine the joint state functions $|{\bf f}\rangle $ of the system $\mathcal{S}$. 
	\item $L_{4}:$ Apply all possible logic gates (differential operators)  between the subsystems $\{\mathcal{S}_{i}\}$, i.e.  ${\bf L}|{\bf f}\rangle$. 
	\item $L_{5}:$  List all  produced patterns and classify them into classical and quantum patterns. 
	\item $L_{6}:$ Compute the expectation values i.e. $\langle {\bf f}|{\bf L}|{\bf f}\rangle$, with suitable normalization conditions. 
	\item $L_{7}:$ If the joint state function is not written in a closed analytical form but rather in a numerical approximated form. Repeat the previous procedures for other joint state functions at different time iterations.
	\item $L_{8}:$ For sufficient time iterations, list all produced patterns and select the distinguished patterns that correspond to different expectation values in the Bargmann space. 
\end{itemize}
Obtaining such lists of computational patterns for each dynamical system allows for designing novel ways for controlling and manipulating the properties of dynamical systems for efficient computing tasks. Consequently, any program in UPL should be described by the triplet $(\mathfrak{CS},\mathcal{S},\mathcal{L})$, where $\mathfrak{CS}$ is the computation space. For a single-qubit $q_{1}$, $\mathfrak{CS}=\mathcal{H}L^{2}(\mathbb{C}^{2},\mu_{\hbar})$, $\mathcal{S}=\{q_{1}\}$ and $\mathcal{L}$ are the eight instructions defined for the case of a single-qubit.

	 \section{Conclusion}
	 We formulated the theory of computation in terms of holomorphic functions in the Bargmann space. This was done first by writing the quantum logic gates as complex differential operators that act on analytic functions. Classical logic gates are embedded in the quantum gates and may be recovered immediately considering the inputs to be of classical signature (no superposition states are allowed). Moreover, the current formalism enjoys great flexibility since it is connected with the theory of holomorphic  functions and can be applied to any dynamical system.  We presented a recipe for constructing a universal programming language that uses holomorphic functions as logical states and complex differential operators for the transition between different logical states.   We suggest a deep study of the formalism presented here in neural dynamics models such as  Hodgkin–Huxley, FitzHugh–Nagumo, and oscillatory threshold logic based on Josephson junctions and optical oscillators \cite{Koch,neuron,Lynch}. Moreover, building a correspondence between logic gates and the different oscillation states of Belousov–Zhabotinsky reactions can be used for building non-conventional computing devices \cite{Kurmaoto, Adamatzky, Adamatzky1}.
	 \section*{Acknowledgment} The author is very much indebted to Professor Andrew Adamatzky for his comments which helped improving the current  manuscript
  
	 \vskip 5mm
	 
	 {\bf Data availability statement:} Data sharing is not applicable – no new data is generated.
	 \vskip 5mm 
	 
	 {\bf Disclosure statement:} No potential conflict of interest was reported by the author(s).
\section*{Appendix}
Let $U$ be a non-empty open set in $\mathbb{C}^{d}$ and let $\mathcal{H}(U)$ denote the space of holomorphic functions on $U$. A complex-valued function $f: U\rightarrow \mathbb{C}$ is said to be holomorphic if it is differentiable in a neighborhood of each point in $U$.    Let $\mathcal{H}L^{2}(U,\omega)$ denotes the space of $L^{2}$-holomorphic functions with respect to the weight $\omega$, that is, 
	 \begin{eqnarray}
	 	\mathcal{H}L^{2}(U,\omega)= \{ F\in \mathcal{H}(U) | \int_{U} |F(z)|^{2}\;\omega(z)\; dz< \infty\}. 
	 \end{eqnarray}
 Here $dz$ denotes the $2d$-dimensional Lebesgue measure on $\mathbb{C}^{d}=\mathbb{R}^{2d}$ not a line integral. Some examples of holomorphic function spaces  are the weighted Bergman  and the  Segal-Bargmann spaces \cite{Hall}. In this work, we used the Segal-Bargmann spaces .  \vskip 5mm
The  Segal-Bargmann spaces (also known as Bargmann spaces) $\mathcal{H}L^{2}(\mathbb{C}^{d},\mu_{t})$	 are spaces 	of holomorphic functions with Gaussian integration measure $\mu_{t}=(\pi)^{-d}e^{-|z|^{2}/t}$
	and  inner product of the form \cite{Hall}
\begin{eqnarray}
	\langle f|g\rangle=(\pi t)^{-d}\int_{\mathbb{C}^{d}}\overline{f}(z)\;g (z)\; e^{-|z|^{2}/t}dz, 
\end{eqnarray}
where $|z|^{2}=|z_{1}|^{2}+\dots +|z_{d}|^{2}$ and $dz$ is the $2d$-dimensional Lebesgue measure on $\mathbb{C}^{d}$ and $t$ is a constant equals to  $\hbar$ if the physical states under study are quantized.\vskip 5mm  
The Segal-Bargmann transform is a  unitary map $B_{t}:L^{2}(\mathbb{R}^{d},dx)\rightarrow \mathcal{H}L^{2}(\mathbb{C}^{d},\mu_{t})$ defined as 
\begin{eqnarray}
	B_{t}f(z)=(\pi t)^{-d/4}\int_{\mathbb{R}^{d} }e^{\left(-z^{2}+2\sqrt{2}z\cdot x-x^{2}\right)/2t} f(x)\;dx.
\end{eqnarray}

The  reproducing kernel  is 
\begin{eqnarray}
	K(z,w)=\sum_{n=0}^{\infty}\frac{z^{n}}{\sqrt{n!t^{n}}}\frac{\overline{w}^{n}}{\sqrt{n!t^{n}}}=\sum_{n=0}^{\infty}\frac{1}{n!}(\frac{z\overline{w}}{t})^{n}=e^{z\overline{w}/t}. 
\end{eqnarray}
 The kernel   $K(z,w)$ is holomorphic in $z$ and anti-holomorphic in $w$. It  satisfies the following properties: \vskip 5mm
 1- $K(z,w)=\overline{K(w,z)}$\vskip 5mm
 2- For each fixed $z\in \mathbb{C}$, $K(z,w)$ is square-integrable $d\mu_{t}$. For all $f(z)\in \mathcal{H}L^{2}(\mathbb{C},\mu_{t})$, we may write 
 \begin{eqnarray}
 	f(z)=\int_{\mathbb{C}}K(z,w)\;f(w) \;\mu_{t}(w)\;dz.
 \end{eqnarray}\vskip 5mm
3- For all $z,w\in \mathbb{C}$ , 
\begin{eqnarray}
	\int_{\mathbb{C}}K(z,u)\;K(u,w)\; \mu_{t}(u) \;du= K(z,w)
\end{eqnarray}
The monomials $\{\frac{z^{n}}{\sqrt{n!t^{n}}}\}$ form an orthonormal basis in $\mathcal{H}L^{2}(\mathbb{C}^{n},\mu_{t})$. The orthonormality condition with $t=d=1$ is 
\begin{eqnarray}
	\frac{1}{\pi }\int_{\mathbb{C}}  e^{-|z|^{2}}\; \overline{z}^{n} z^{m}dz =  n!\; \delta_{mn}.  
\end{eqnarray} 
\vskip 5mm
Using the properties of the Bargamnn spaces, one could calculate the numerical values of the $n$th derivative of any complex functions at point $z_{0}$ by integration over the Bargamnn space. 
Let $f(z)$ be an analytic function of complex variable $z$.  The Taylor series of $f$ at a complex number $z_{0}$ is \cite{Ahlfors}
\begin{equation}\label{taylor}
	f(z)= f(z_{0})+ \frac{f^{\prime}(z_{0})}{1!} \left(z-z_{0}\right)+ \frac{f^{\prime \prime}(z_{0})}{2!} (z-z_{0})^{2}+\dots= \sum_{n=0}^{\infty} \frac{f^{(n)}(z_{0})}{n!} \left(z-z_{0}\right)^{n}.
\end{equation}
When $z_{0}=0$, the Taylor series  reduces to a Maclaurin series i.e. $\sum_{n=0}^{\infty} \frac{f^{(n)}(0)}{n!} z^{n} $.  Let us compute the inner product $\langle z^{n}|f(z)\rangle$, where $f(z)=  \sum_{n=0}^{\infty} \frac{f^{(n)}(0)}{n!} z^{n}$ is a Maclaurin series, we find 
\begin{eqnarray}\label{result}
	f^{(n)}(0)=  \langle z^{n}|f(z)\rangle, 
\end{eqnarray} 
which is very interesting since it gives a closed analytical expression for the $n$th derivative of any analytic complex function at $z=0$ by computing the inner product of its Maclaurin series expansion with monomial $z^{n}$  in $\mathcal{H}L^{2}(\mathbb{C},\mu_{1})$.  Practically, this  means computing Gaussian integrals of the form $\frac{1}{\pi}\int_{\mathbb{C}} \overline{z}^{n} f(z)e^{-|z|^{2}} dz$. For the Taylor series, the formula \ref{result} becomes
\begin{eqnarray}\label{resulttaylor}
	f^{(n)}(z_{0})= \langle (z-z_{0})^{n}|f(z)\rangle.
\end{eqnarray}

The derivative of a given analytic function $f(z)$ is 
\begin{eqnarray}\label{derivative}
	f^{\prime}(z)= \lim_{h\rightarrow 0}\frac{f(z+h)-f(z)}{h}, 
\end{eqnarray}
where $h$ is the step size. For infinitesimally small $h$, we may compute the derivative approximately using a forward-difference expression, 
\begin{eqnarray}\label{forward}
	f^{\prime}(z)\approx \frac{f(z+h)-f(z)}{h},
\end{eqnarray}
Using \ref{resulttaylor}, the forward-difference expression at point $z_{0}$ assumes the form 
\begin{eqnarray}\label{result1}
	\langle (z-z_{0})|f(z)\rangle \approx \frac{1}{h}\left(f(z_{0}+h)-f(z_{0})\right)
\end{eqnarray}
Analogously, the second-order forward-difference expression is 
\begin{eqnarray}
	\langle (z-z_{0})^{2}|f(z)\rangle \approx  \frac{1}{h^{2}}\left(f(z_{0}+2h)-2f(z_{0}+h)+f(z_{0})\right). 
\end{eqnarray}
Thus, we have written the finite differences in terms of integrals over the space of holomorphic function spaces with Gaussian integration measure. Consequently,  one could use the numerical methods for integrals to compute derivatives approximately instead of finite differences\cite{Hamming}. This is more efficient since for higher-order derivatives we would have long expressions using the finite-difference method. However, using our method we can compute higher-order derivatives using one integral  similar to \ref{result}.
\vskip 5mm
Now let us look closely at \ref{result1}, using the fundamental theorem of calculus we find 
\begin{eqnarray}
	f^{\prime}(z_{0})=\langle (z-z_{0})|f(z)\rangle\approx \frac{1}{h}\int_{z_{0}}^{z_{0}+h} f^{\prime}(z) dz. 
\end{eqnarray}
Interestingly, the derivative of function $f$ at point $z_{0}$   is approximately equal to the integral of $f^{\prime}(z)$ between $z_{0}+h$ and $z_{0}$. This means that we can equate integrals over holomorphic function with Gaussian integration measure with integrals over ordinary complex space.\vskip 5mm

The properties of inner products in the Bargmann spaces can be used in computing the coefficients of series expansions due to the orthonormality of holomorphic variables in the Bargmann spaces.  As an  example, the generalized hypergeometric function is 
\begin{eqnarray}
	_{p}F_{q}\left(a_{1},\dots, a_{p}; b_{1},\dots,b_{q};z\right)=\sum_{n=0}^{\infty}\frac{(a_{1})_{n}\dots (a_{p})_{n}}{(b_{1})_{n}\dots (b_{q})_{n}}\frac{z^{n}}{n!},
\end{eqnarray}
where $(x)_{n}=x(x+1)(x+2)\dots (x+n-1), \; \; n\geq 1$ is the  Pochhammer symbol with $(x)_{0}=1$. Using  \ref{result}, we find 
\begin{eqnarray}
	\frac{(a_{1})_{n}\dots (a_{p})_{n}}{(b_{1})_{n}\dots (b_{q})_{n}}= \langle z^{n}|	_{p}F_{q}\left(a_{1},\dots, a_{p}; b_{1},\dots,b_{q};z\right)\rangle .
\end{eqnarray}
Thus, the coefficients of generalized hypergeometric functions can be determined uniquely using inner products in  Bargmann spaces. 
	
\end{document}